\documentclass[prb,aps,twocolumn,showpacs,nofootinbib]{revtex4}
\usepackage[dvips]{graphicx}
\usepackage{amsmath}
\usepackage{epsfig}
\newcommand{\bea}{\begin{eqnarray}}
\newcommand{\eea}{\end{eqnarray}}
\newcommand{\be}{\begin{equation}}
\newcommand{\ee}{\end{equation}}

\newcommand{\BSCCO}{{Bi$_2$Sr$_2$CaCu$_2$O$_8$ }}

\newcommand{\LSCO}{{La$_{1-x}$Sr$_x$CuO$_4$}}
\begin{document}


\title{Magnetic correlations on the full chains of Ortho-II YBa$_2$Cu$_3$O$_{6.5}$}
\author{W. Chen$^{1}$ and P.J. Hirschfeld$^{2}$}
\affiliation{
$^{1}$School of Physics, University of New South Wales, Kensington 2052, Sydney NSW, Australia \\
$^{2}$Department of Physics, University of Florida, Gainesville,
FL 32611 USA }
\begin{abstract}
We propose that the NMR line shape on the chain Cu in the
stoichiometric high-$T_c$ superconductor Ortho-II
YBa$_2$Cu$_3$O$_{6.5}$ is determined by the magnetization induced
on  Cu near O vacancies, due to strong magnetic correlations in
the chains.  An unrestricted Hartree-Fock calculation of a coupled
chain-plane Hubbard model with nearest-neighbor $d$-wave pairing
interaction shows that the broadening of NMR lines is consistent
with disorder-induced magnetization at low temperatures. In
addition, we give a possible explanation of the anomalous bimodal
line shape observed at high temperatures in terms of nonuniform Cu
valence  in the chains. The proximity between chains and CuO plane
induces anisotropic magnetization on the planar Cu, and broadens
the plane NMR lines in accordance with that of the chain lines, in
agreement with experiment.  We discuss implications of the model
for other experiments on underdoped YBCO. \vskip .4cm
\begin{center}(\today)\end{center}
\end{abstract}

 \maketitle

\section{Introduction}


In  materials with strong electronic correlations like the
cuprates, the response of the system to  disorder can be quite
different from weakly interacting metals. One well-known anomalous
effect observed in disordered cuprates is the extended staggered
magnetic droplets carrying a net moment which form  around
nonmagnetic impurities introduced on the CuO$_{2}$ plane.  This
picture was proposed on the basis of extensive  NMR studies, and
can be understood within various theories of a potential scatterer
in a correlated host metal\cite{Alloul07}. The local
susceptibility of these magnetic droplets manifests a Curie-Weiss
behavior with a Weiss temperature which drops rapidly as the
materials are underdoped\cite{Bobroff01}. While this picture was
established for Zn, Li, and strong scatterers in the CuO$_2$ plane
created by electron irradiation, the magnetic response to the
weaker potentials created by out-of-plane dopants is also expected
to be enhanced, and has been claimed to be responsible for glassy
behavior observed at low $T$ and doping in intrinsically
disordered systems such as \LSCO and \BSCCO\cite{Andersen07}. In
such materials, it is generally believed that quasiparticles
moving in the CuO$_2$ plane are subjected to weak, extended
potentials caused by the dopants, giving rise to  small momentum
transfer scattering in the electronic
transport\cite{Kee,Abrahams,Zhu04}.

In the YBa$_{2}$Cu$_{3}$O$_{7-\delta}$(YBCO) family where most of
the impurity induced magnetization effects were studied, the
system is doped in a different manner, which depends on the
distribution of O atoms in the chain layers.  Careful annealing
under uniaxial pressure has been shown to give high quality single
crystals where the chain O is ordered with long correlation
lengths\cite{Liang99}. The so-called Ortho-I, Ortho-II, etc. YBCO
crystals formed by these methods are the only known doped,
ordered, stoichiometric high-$T_c$ materials, ideal for studying
the underlying intrinsic physics of the cuprates without
complications from dopant disorder.   It is then interesting to
explore the effects of the few isolated defects which remain in
these crystals,  expected to be O vacancies in otherwise full CuO
chains\cite{Bobowski07}.

The one dimensional nature of the chains themselves induces, in
the filled chain compound, $a-b$ anisotropy which has been
observed in dc\cite{Friedmann90,Gagnon94} and optical
conductivity,\cite{Basov95} penetration depth
measurements,\cite{Zhang94,Tallon95} and is expected to affect the
vortex core structure\cite{Atkinson08}. Several authors have
proposed that the anisotropy in penetration depth can be explained
by assuming a metallic chain that couples to the plane via
interlayer
hopping\cite{Atkinson95,Atkinson97,ODonovan97,Combescot98,Morr01}.
The superconductivity observed in the chains is assumed to be due
to the proximity coupling to the plane, where pairing occurs.
While initially quantitative details of penetration depth
disagreed with
the proximity models, 
Atkinson has argued that accounting for disorder can remove  these
discrepancies\cite{Atkinson99}.

The presence of disorder has been shown to modify the local
electronic structure of the chain layer. Scanning tunnelling
spectroscopy (STS) measurements on the chain layer of optimally
doped  YBCO show a clear conductance
modulation,\cite{Edwards92,Edwards94,Edwards95} whose wavelength
displays a strong energy dependence.\cite{Derro02} Such a
dispersion in a quantity directly related to the local density of
states (LDOS) suggests a Friedel-type oscillation description of
this phenomenon, in contrast to an explanation in terms of a
charge density wave(CDW), which usually has a fixed wave vector.
In addition, resonance peaks that appear in pairs are found at
small frequencies, reminiscent of   resonant magnetic impurity
states in an unconventional superconductor.   Morr and Balatsky
\cite{Morr03} argued that such features could be caused by
magnetic impurities in the chain layer, without discussing the
origin of such defects.

In this paper we study the YBCO compound with a particular
out-of-plane oxygen distribution, namely the Ortho-II structure of
YBCO6.5 where every other chain is filled, and show that the
existence of magnetic correlations and their interplay with
defects on the chain can explain a series of NMR experiments by
Yamani $et\;al$\cite{Yamani03,Yamani06}. We show in Section II
that the magnetization resulting  from the  Friedel oscillations
of an uncorrelated system near the chain ends  gives NMR
linewidths much smaller than what has been reported.  The line
width in the uncorrelated case also remains
temperature-independent, contradicting the experimentally observed
broadening with decreasing temperature. Including magnetic
correlations in the chain layer within  weak coupling mean field
theory, we show that both the significantly larger linewidth and
temperature dependence
 can be accounted for. This  demonstrates the importance of magnetic correlations
in the chains of YBCO systems, which is frequently neglected.

One of the unusual features of the lineshape observed on the chain
Cu's  is a low-frequency satellite peak.
We show here in Section III that satellite features do  not arise
naturally due to O chain vacancies, but must be related to a set
of sites in the material which have zero spin shift. One
possibility we discuss in some detail is that some of the chain
Cu's are in the Cu$^3+$ configuration.

 In Section IV, we consider the coupling of the chain to the CuO$_2$ plane, and show
that the major effect of the chain-plane coupling is to induce an
anisotropic magnetization pattern on the plane, where the chains imprints
their 1D correlation length onto the plane. The broadening of plane NMR lines in
accordance with that of chain NMR lines is reproduced in the
proposed model, consistent with what is observed in the
correlation between chain and plane linewidth in experiment.


\section{Model Hamiltonian and  NMR Spectrum}

The proper choice of a model for the Ortho-II system relies on its
unique lattice structure.  Due to missing oxygens on the empty
chain, the Cu(1E)(chain Cu in ``empty" chain) sites are highly
localized and do not directly couple to the full chain; we
therefore assume they do not affect the magnetic properties on the
full chain and drop these degrees of freedom. The resulting
minimum effective model contains a single square lattice of
Cu(2E)(planar Cu above empty chain site) and Cu(2F)(planar Cu
above full chain site) where superconductivity occurs, couples to
evenly spaced one dimensional chains via interlayer hopping, as
shown in Fig. \ref{fig:YBCO_orthoII_lattice}. The full Hamiltonian
consists of

\begin{equation}
H=H_{p}+H_{c}+H_{inter}+H_{imp}\;,
\label{chain_bare_H}
\end{equation}
where $H_{p}$ describes the planar hopping, pairing and magnetic correlations

\begin{eqnarray}
H_{p}&=&\sum_{ij\sigma}-t^{p}_{ij}{\hat c}^{\dag}_{i\sigma}{\hat c}_{j\sigma}
+\sum_{i\sigma}(\epsilon^{p}_{\sigma}-\mu^{p}){\hat n}^{p}_{i\sigma}\\
&+&\sum_{i}U^{p}{\hat n}^{p}_{i\uparrow}{\hat n}^{p}_{i\downarrow}
-\sum_{\langle ij\rangle}V{\hat n}^{p}_{i\uparrow}{\hat n}^{p}_{j\downarrow}\;,
\end{eqnarray}
and on the chain layer

\begin{eqnarray}
H_{c}&=&\sum_{x\sigma}-t^{c}({\hat d}^{\dag}_{x\sigma}{\hat d}_{x+\delta\sigma}+h.c.)\\
&+&\sum_{x\sigma}(\epsilon^{c}_{\sigma}-\mu^{c}){\hat
n}^{c}_{x\sigma} +\sum_{x}U^{c}{\hat n}^{c}_{x\uparrow}{\hat
n}^{c}_{x\downarrow}\;,
\end{eqnarray}
with the coupling between them

\begin{equation}
H_{inter}=\sum_{\langle ix\rangle\sigma}-t^{r}({\hat c}^{\dag}_{i\sigma}{\hat d}_{x\sigma}+h.c.)\;.
\end{equation}
We denote by ${\hat c}_{i\sigma}$(${\hat d}_{x\sigma}$) the
plane(chain) operator located at site i(x), and
$t^{p}_{ij}$($t^c$) the planar(chain) hopping, with on site
Coulomb repulsion $U^p$($U^c$) on the plane(chain). The chain
hopping is only between nearest neighbors,  while planar hopping
$t^{p}_{ij}$ contains both nearest $t^{p}_{nn}$ and next nearest
neighbor sites $t^{p}_{nnn}$ which is necessary to produce the
Fermi surface of the YBCO system. The short range attractive
interaction $V$ accounts for the $d$-wave pairing in the plane,
and the effect of magnetic field is included in the Zeeman term
$\epsilon^{p/c}_{\sigma}=g\mu_{B}B/2$. The proper choice of all
parameters except the $U$'s relies on three criteria: (1)the Fermi
surface of ortho-II calculated from  density functional
theory(DFT)\cite{Bascones05,Carrington07} is reproduced in this
simple 3-site model; (2)the homogeneous magnetization on the full
chains is consistent with the value indicated by the main line of
Cu(1F) NMR spectrum, and (3) a critical temperature of $T_{c}=60K$
is obtained. Regarding the first point, Fig.
\ref{fig:YBCO_orthoII_lattice} shows the Fermi surface produced by
the present model in the normal state with $V=0$. Although one
does not expect the splitting of the plane band since the chains
are connected to a single plane in this minimum model, the
topology and the folding of the Brillouin zone is well reproduced.
Notice that the chain band is almost parallel to the zone boundary
with a very small intercept $k_{y}\sim \pi/4$, indicating the
small filling of the conduction electrons on the chain, which is a
natural result of fitting the complex Fermi surface by this simple
tight binding model. We therefore choose the chemical potential
such that the average chain filling is $0.25$, and the average
plane filling is  $0.9$. Since the main line of the NMR spectrum
is given by the sites far from the defects if one assumes a dilute
impurity concentration, we found that $t^{c}=2$ gives a Knight
shift and magnetization consistent with the value corresponding to
the measured Cu(1F) main line, which then fixes the absolute scale
of the band width for the chain. Finally, $T_{c}=60K$ in the
presence of the chosen interlayer hopping fixes the pairing
interaction at $V=0.5$, and the absolute scale of our energy unit $t^{p}_{nn}$
is chosen such that the magnetic field scale and temperature dependence of
the NMR lines are consistent with experiments, which gives $t^{p}_{nn}=150$meV.

\begin{figure}[ht]
\begin{center}
\leavevmode
\includegraphics[clip=true,width=0.8\columnwidth]{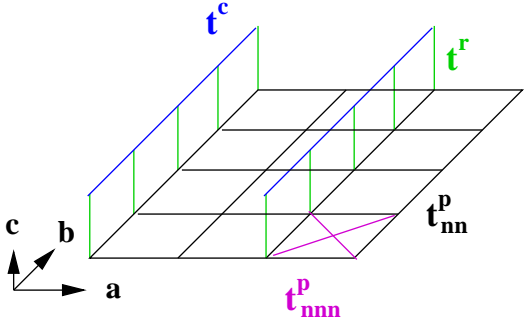}
\caption{Schematic of chain-plane proximity model for Ortho-II
YBCO.}\label{fig:hoppings_schematic}
\end{center}
\end{figure}

A Hartree-Fock-Gorkov mean field factorization is then applied to
the above Hamiltonian,  with order parameters defined as
\begin{eqnarray}
n^{p}_{i}&=&\langle {\hat n}^{p}_{i\uparrow}+{\hat n}^{p}_{i\downarrow}\rangle\;,\nonumber\\
m^{p}_{i}&=&\langle {\hat n}^{p}_{i\uparrow}-{\hat n}^{p}_{i\downarrow}\rangle\;,\nonumber\\
\Delta_{\delta i}&=&
V\langle c_{i\uparrow}c_{i+\delta\downarrow} \rangle\;,\nonumber\\
n^{c}_{x}&=&\langle {\hat n}^{c}_{x\uparrow}+{\hat n}^{c}_{x\downarrow}\rangle\;,\nonumber\\
m^{c}_{x}&=&\langle {\hat n}^{c}_{x\uparrow}-{\hat n}^{c}_{x\downarrow}\rangle\;.
\label{chain_op}
\end{eqnarray}
A Bogoliubov transformation is applied for both chain and plane operators
\begin{eqnarray}
c_{i\uparrow}&=&\sum_{n}u^{p}_{n,i\uparrow}\gamma_{n\uparrow}
+v^{p\ast}_{n,i\uparrow}\gamma^{\dag}_{n\uparrow} \nonumber\\
c_{i\downarrow}&=&\sum_{n}u^{p}_{n,i\downarrow}\gamma_{n\downarrow}
+v^{p\ast}_{n,i\downarrow}\gamma^{\dag}_{n\downarrow} \nonumber\\
d_{x\uparrow}&=&\sum_{n}u^{c}_{n,x\uparrow}\gamma_{n\uparrow}
+v^{c\ast}_{n,x\uparrow}\gamma^{\dag}_{n\uparrow} \nonumber\\
d_{x\downarrow}&=&\sum_{n}u^{c}_{n,x\downarrow}\gamma_{n\downarrow}
+v^{c\ast}_{n,x\downarrow}\gamma^{\dag}_{n\downarrow}\;,
\label{chain_Bogoliubov}
\end{eqnarray}
and the order parameters are determined by the following self-consistent equations
\begin{eqnarray}
\langle c^{\dag}_{i\uparrow}c_{i\uparrow}\rangle
&=&\sum_{n>0}|u^{p}_{n,i\uparrow}|^{2}f(E_{n\uparrow})
+|v^{p}_{n,i\uparrow}|^{2}(1-f(E_{n\downarrow}))\;,
\nonumber\\
\langle c^{\dag}_{i\downarrow}c_{i\downarrow}\rangle
&=&\sum_{n>0}|u^{p}_{n,i\downarrow}|^{2}f(E_{n\downarrow})
+|v^{p}_{n,i\downarrow}|^{2}(1-f(E_{n\uparrow}))\;,
\nonumber\\
\langle d^{\dag}_{x\uparrow}d_{x\uparrow}\rangle
&=&\sum_{n>0}|u^{c}_{n,x\uparrow}|^{2}f(E_{n\uparrow})
+|v^{c}_{n,x\uparrow}|^{2}(1-f(E_{n\downarrow}))\;,
\nonumber\\
\langle d^{\dag}_{x\downarrow}d_{x\downarrow}\rangle
&=&\sum_{n>0}|u^{c}_{n,x\downarrow}|^{2}f(E_{n\downarrow})
+|v^{c}_{n,x\downarrow}|^{2}(1-f(E_{n\uparrow}))\;,
\nonumber\\
\langle c_{i\uparrow}c_{i+\delta\downarrow}\rangle
&=&\sum_{n>0}u^{p}_{n,i\uparrow}v^{p\ast}_{n,i+\delta\downarrow}(1-f(E_{n\uparrow}))
\nonumber\\
&+&v^{p\ast}_{n,i\uparrow}u^{p}_{n,i+\delta\downarrow}f(E_{n\downarrow})\;,
\nonumber\\
\langle d_{x\uparrow}d_{x+\delta\downarrow}\rangle
&=&\sum_{n>0}u^{c}_{n,x\uparrow}v^{c\ast}_{n,x+\delta\downarrow}(1-f(E_{n\uparrow}))
\nonumber\\
&+&v^{c\ast}_{n,x\uparrow}u^{c}_{n,x+\delta\downarrow}f(E_{n\downarrow})\;,
\end{eqnarray}
where $n>0$ indicates only eigenstates with positive eigenenergies
are included in the  summation. Note that by estimating the
concentration of terminal Cu, the average chain length that
produces reported NMR spectrum was reported to be around $120b$,
where $ b$ is the chain lattice constant by Yamani et al.   We
therefore simulate such a system by a rectangular plane with size
$12\times 121$, where each odd chain allows for hopping along its
 length of $121$ sites directed along the x-direction.

\begin{figure}[t]
\begin{center}
\includegraphics[clip=true,width=0.8\columnwidth]{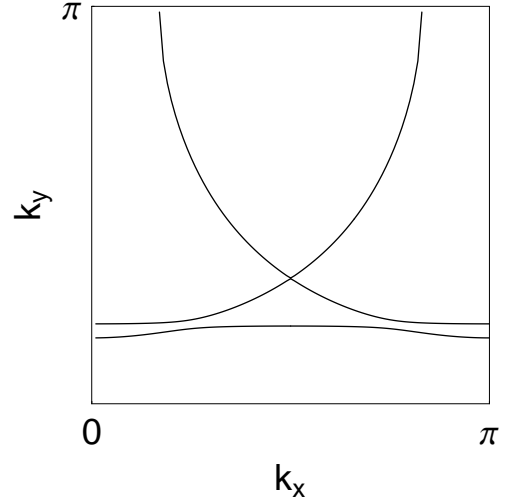}

\caption{(color online).  Fermi surface produced by model
described in Fig. \ref{fig:hoppings_schematic} with
parameters$(t^{p}_{nn}=1,t^{p}_{nnn}=-0.4,t^{c}=2,t^{r}=0.1,\mu^{p}=-1.1,\mu^{c}=-3.25)$,
and we take $t^{p}_{nn}=150$meV.} \label{fig:YBCO_orthoII_lattice}
\end{center}
\end{figure}

The resonance frequency $\nu_{i}$ at site $i$ is calculated by
converting magnetization to Knight shift $K_{spin}$,
\begin{eqnarray}
S_{x}&=&\frac{1}{2}\langle n_{x\uparrow}-n_{x\downarrow}\rangle=\frac{\chi_{x}B}{g\mu_{B}}\;,\nonumber\\
\chi_{x}&=&\frac{K_{spin}\mu_{B}}{A_{hf}}\;,
\end{eqnarray}
which yields, when combined with the orbital shift $K_{orb}$, the
local resonance frequency
\begin{equation}
\nu_{x}=\frac{\gamma}{2\pi}B[1+K_{orb}+S_{x}(\frac{gA_{hf}}{B})]\;.
\label{chain_NMR_formula}
\end{equation}
For Cu(1F) NMR lines, we use $A_{hf}=80$kOe, $g=2$,
$K_{orb}=1.2\%$, and  $\gamma/2\pi=11.285$MHz/T. The histogram of
collecting $\nu_{i}$ on all Cu(1F) sites is then artificially
broadened by a Lorentzian to give a continuous spectrum
\begin{equation}
I(\nu)=\frac{1}{R}\sum_{\nu_{x}}N(\nu_{x})\frac{1}{\pi}\frac{\eta}{(\nu-\nu_{x})^{2}+\eta^{2}}\;,
\label{chain_LZ_arti_broaden}
\end{equation}
where $N(\nu_{i})$ is number of sites that have frequency
$\nu_{i}$, with $\eta=0.04$ and $R$  is the proper numerical
factor that normalizes the area under the $I(\nu)$ curve. 

\begin{figure}[ht]
\centering
\begin{tabular}{cccc}
  \epsfig{file=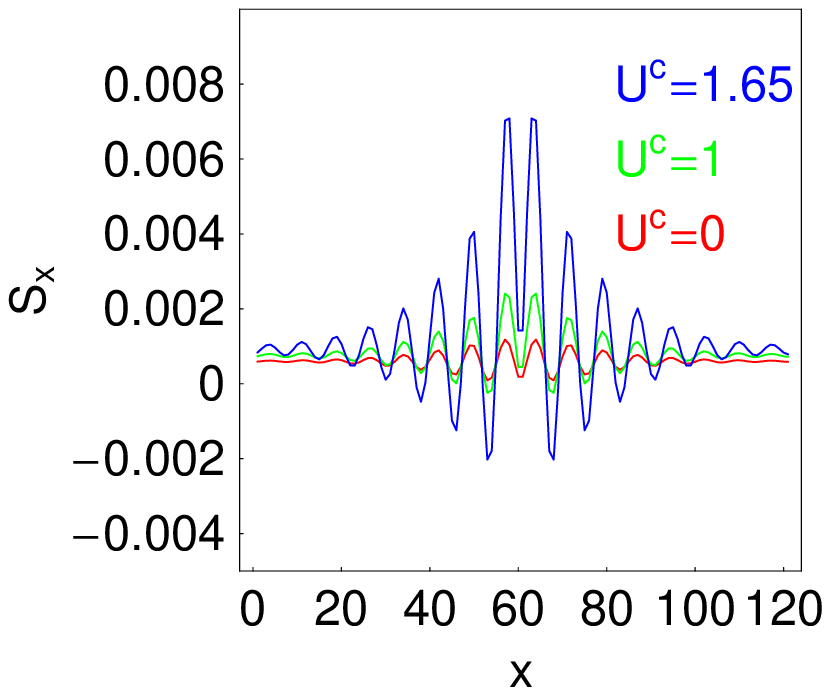,width=4.5cm,clip=true}&
  \epsfig{file=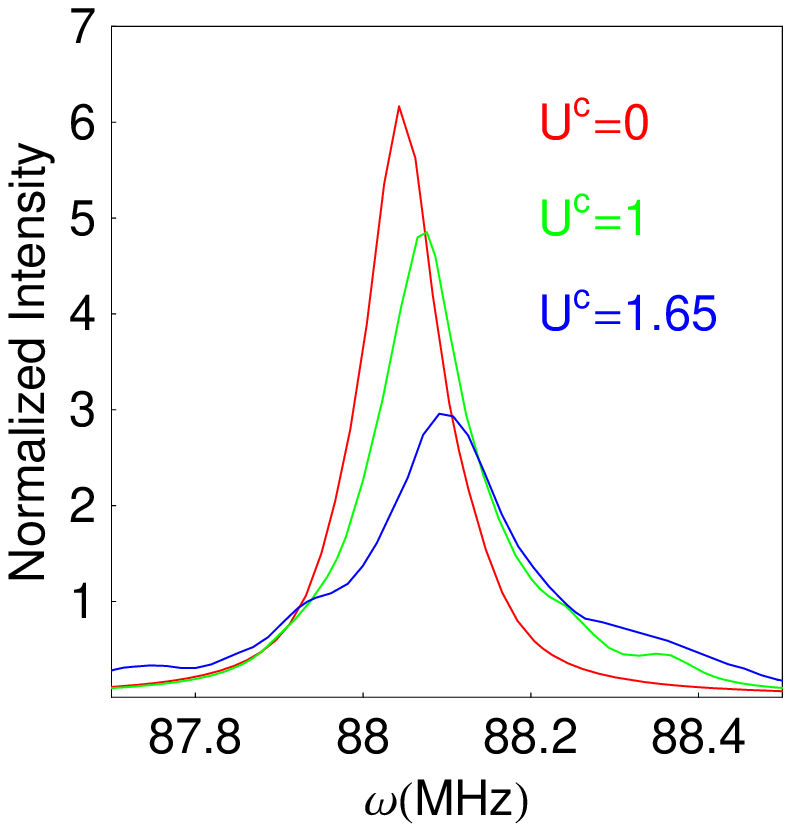,width=4cm,clip=true}&
\end{tabular}
\caption{(color online) Magnetization(left) and NMR spectrum(right)
produced by reduction of a single  hopping in an isolated 1D chain
(located at the middle of the chain as shown), equivalent to
Hamiltonian $H_{chain}+H^{(1)}_{imp}$, in an external field 7.7T
at 50K. The enhancement of magnetization is consistent with the
broadening of NMR linewidth as $U^{c}$ is increased.}
\label{fig:bondimp_scanU_50K_NMR}
\end{figure}

We now examine different impurity models that are relevant to the
NMR experiments. The most straightforward proposal for $H_{imp}$
comes from the observation that the chain hopping is a result of
orbital overlap between Cu(1F) and its two adjacent oxygens, such
that  abrupt termination of the full chain due to missing oxygens
should dramatically reduce the hopping at the chain ends.  A
single isolated reduction of $t^{c}$ is modelled by an additional
term in the Hamiltonian
\begin{equation}
H^{(1)}_{imp}=\sum_{\sigma}-\delta t^{c}(d^{\dag}_{l\sigma}d_{l+\delta\sigma}+h.c.)\;,
\label{chain_1D_imp_H}
\end{equation}
with a further assumption of complete elimination $\delta
t^{c}=-t^{c}$ and one impurity per chain. The Cu(1F) line caused
by a single hopping reduction in the chain layer is shown in Fig.
\ref{fig:bondimp_scanU_50K_NMR}, where we highlight the effect of
magnetic correlation $U^{c}$ with the model containing only
$H_{chain}+H^{(1)}_{imp}$, to isolate the chain from proximity to
the plane in the first analysis.

\begin{figure}[t]
\centering
\begin{tabular}{cccc}
\epsfig{file=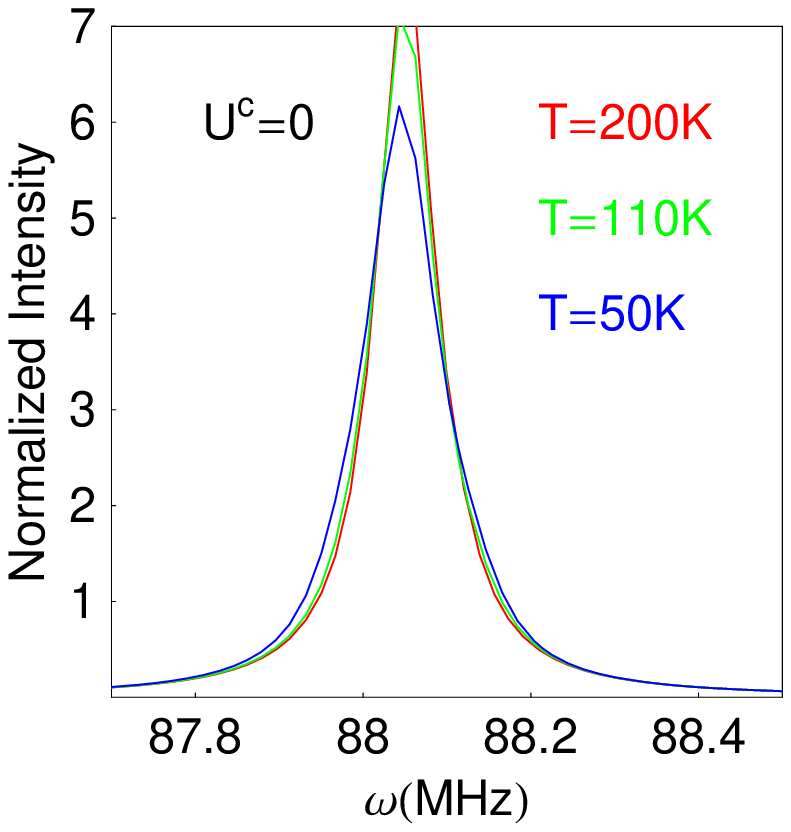,width=4cm,clip=true} &
\epsfig{file=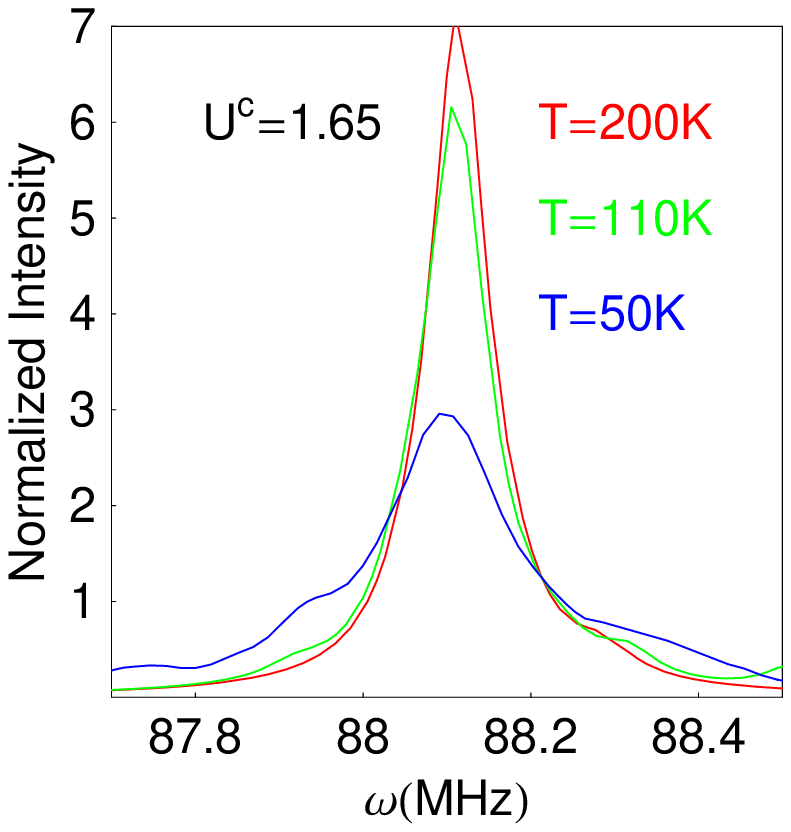,width=4cm,clip=true} &
\end{tabular}
\caption{(color online) Temperature dependence of an NMR lines for
an isolated chain with sparse bond impurities,
$H_{chain}+H^{(1)}_{imp}$: for an uncorrelated chain(left), and
the correlated case(right).} \label{fig:chain_correlated_NMR_scanT}
\end{figure}

The parameter range explored for $U^{c}$ is such that it is close
but smaller than the critical value beyond which the system flows
into long range magnetic order in the presence of impurities, and
the induced moment remains in the paramagnetic region. Fig.
\ref{fig:bondimp_scanU_50K_NMR} shows that the broadening of the
Cu(1F) line coincides with the enhancement of real space
magnetization as $U^{c}$ is increased, while the linewidth for the
uncorrelated case $U^{c}=0$ is significantly smaller.   The
periodicity of magnetic oscillation, as well as the density
modulation for electrons of both spin species, is of the order of
$10$ lattice constants from the defect, consistent with the small
Fermi momentum of the chain band $k_{y}\sim\pi/4$ resulting from
fitting the Fermi surface with the 3-site model. Moreover, the
temperature dependence of the linewidth shown in Fig.
\ref{fig:chain_correlated_NMR_scanT} indicates that only the
correlated case can properly account for the experimentally
observed  broadening as temperature is lowered, while the width in
the uncorrelated case remains constant. The linewidth plus its
temperature dependence therefore proves the existence of magnetic
correlations on the full chains of the ortho-II YBCO6.5. Our best
fit to the line width gives $U^{c}=1.65$, and will be the value
used for the rest of the paper unless otherwise specified.  We
make detailed comparison with experiment in Section IV.

\section{Satellite feature in chain C\lowercase{u} NMR line}

Although the Cu(1F) linewidth and temperature dependence is well
described by impurity induced magnetization due to $U^{c}$, one
important feature regarding the Cu(1F) line shape seems to be
outside of this scheme. At temperature higher than $70K$ or so, a
satellite peak with significant weight gradually develops at
frequency slightly lower than the main line\cite{Yamani06}.
Our calculations for a single defect or chain end  indicate that
magnetization decays smoothly
 away from the chain ends, however, leading to no particular satellite feature in the
 histogram of a single defect. We therefore propose an alternate explanation for the
 satellite peak observed. Two aspects of this high
temperature satellite  appear to us to be important: (1)The weight
of the peak is about $10$\% of the whole spectrum, and (2)the
position of the peak remains roughly the same at all
temperatures. 
Aspect (1) then implies  that roughly $10$\% of the sample has the
same magnetization value, which results in this satellite peak.
Aspect (2) suggests that this magnetization remains constant at
high temperature. Since increasing temperature should continuously
reduce any finite magnetization developed due to the correlation
effect, it is reasonable to assume this observed constant
magnetization is zero. In fact, the frequency of this satellite
peak corresponds very closely to an NMR shift $\nu$ with zero
Knight shift $K_{spin}=0$.   We therefore propose that the high
temperature satellite peak is due to a section of adjacent chain
Cu which have their conduction electrons missing, i.e. Cu$^+$
ions. The valence changes may be correlated with  the chain ends
due to the change of local chemical environment which prohibits
electrons from populating these sites, but need not be. Indeed the
valence of the Cu(1) ions has been controversial, and our analysis
suggests that the valence may be distributed quite
inhomogeneously.
Eliminating these adjacent Cu(1F) Cu in the one band model gives
the following perturbation
\begin{eqnarray}
H^{(2)}_{imp}&=&\sum_{1\leq l\leq L,\sigma}-\delta t^{c}(d^{\dag}_{l\sigma}d_{l+\delta\sigma}+h.c.) \nonumber\\
&+&\sum_{1\leq l\leq L,i,\sigma}-\delta t^{r}({\hat c}^{\dag}_{i\sigma}{\hat d}_{l\sigma}+h.c.)
\nonumber\\
&+&\sum_{1\leq l\leq L,\sigma}U_{imp}{\hat n}^{c}_{x\sigma}\;,
\label{chain_section_imp_H}
\end{eqnarray}
where $\delta t^{c}=-t^{c}$ eliminates the hopping between these
chain sites, and $\delta t^{r}=-t^{r}$ eliminates the interlayer
hopping that connects the plane sites with these chain sites. A
strong impurity potential $U_{imp}=100$ is introduced to
artificially project out electrons on these sites. Notice that the
summation over impurity position in Eq.
(\ref{chain_section_imp_H}) is arbitrarily restricted between a
section of adjacent sites $1\leq l\leq L$ with $L=15$.

\begin{figure}[ht]
\begin{center}
\leavevmode
\begin{minipage}{3.5cm}
\includegraphics[clip=true,width=3.5cm]{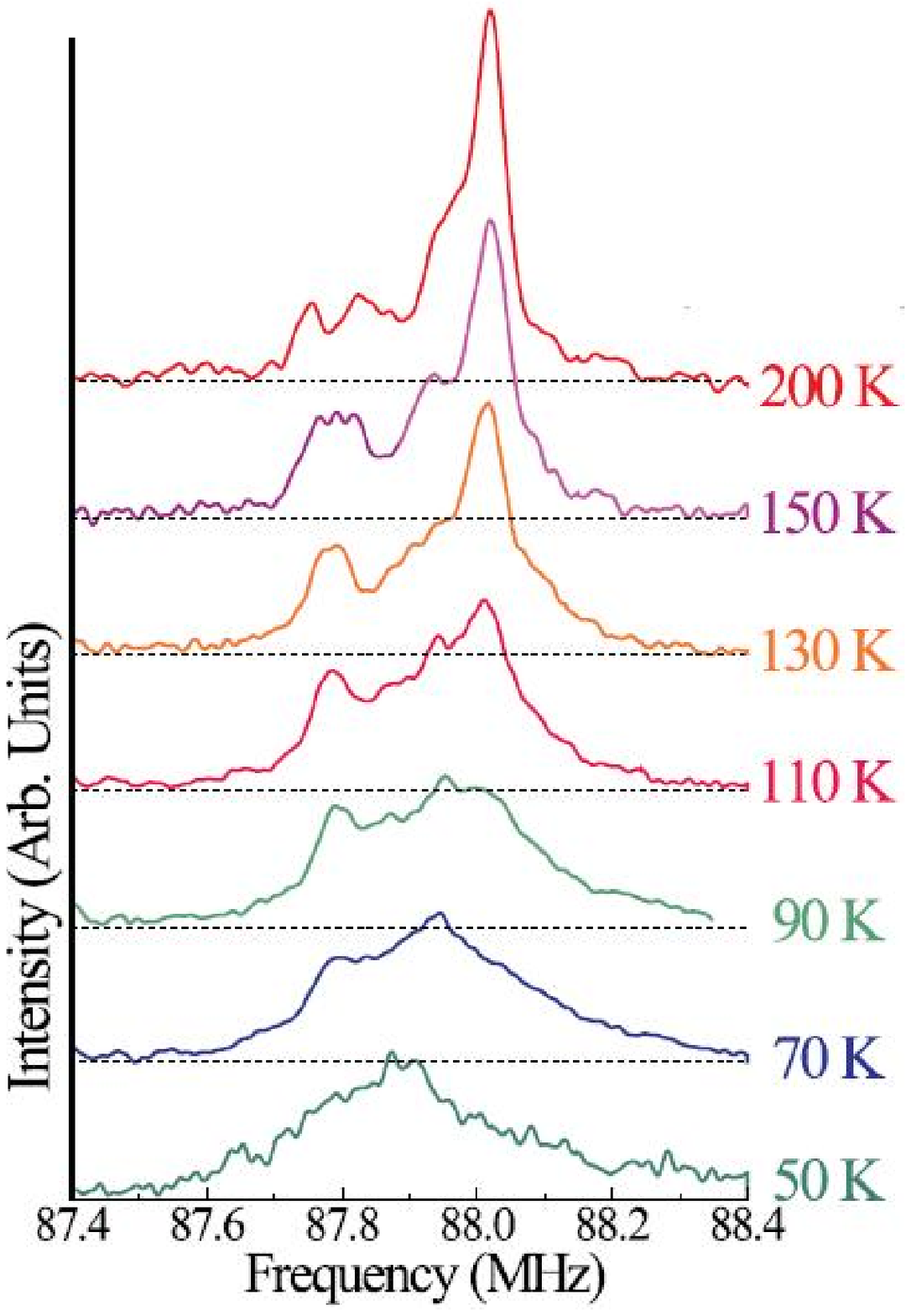}
\end{minipage}
\begin{minipage}{3.5cm}
\includegraphics[clip=true,width=3.5cm]{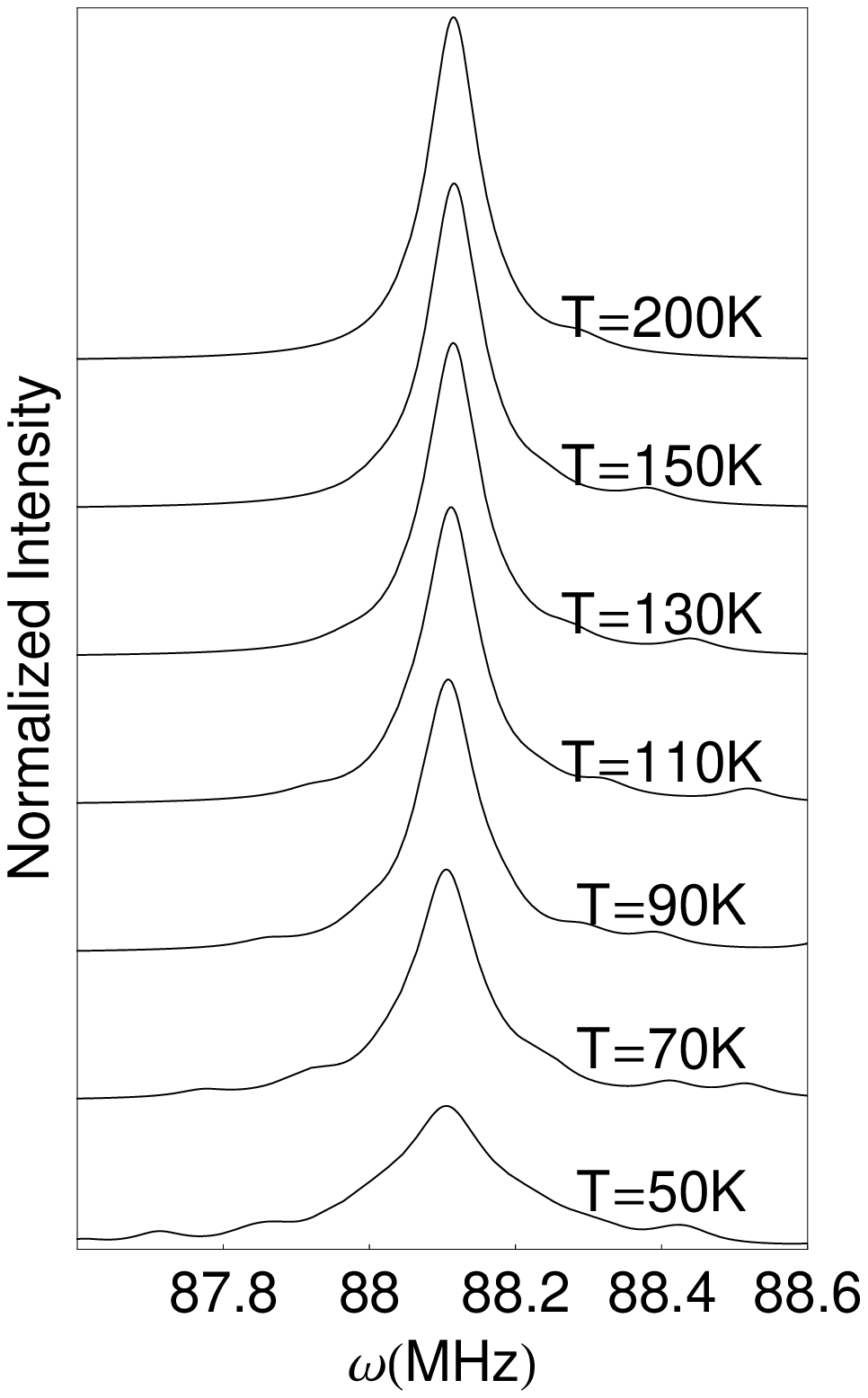}
\end{minipage}
\begin{minipage}{3.5cm}
\includegraphics[clip=true,width=3.5cm]{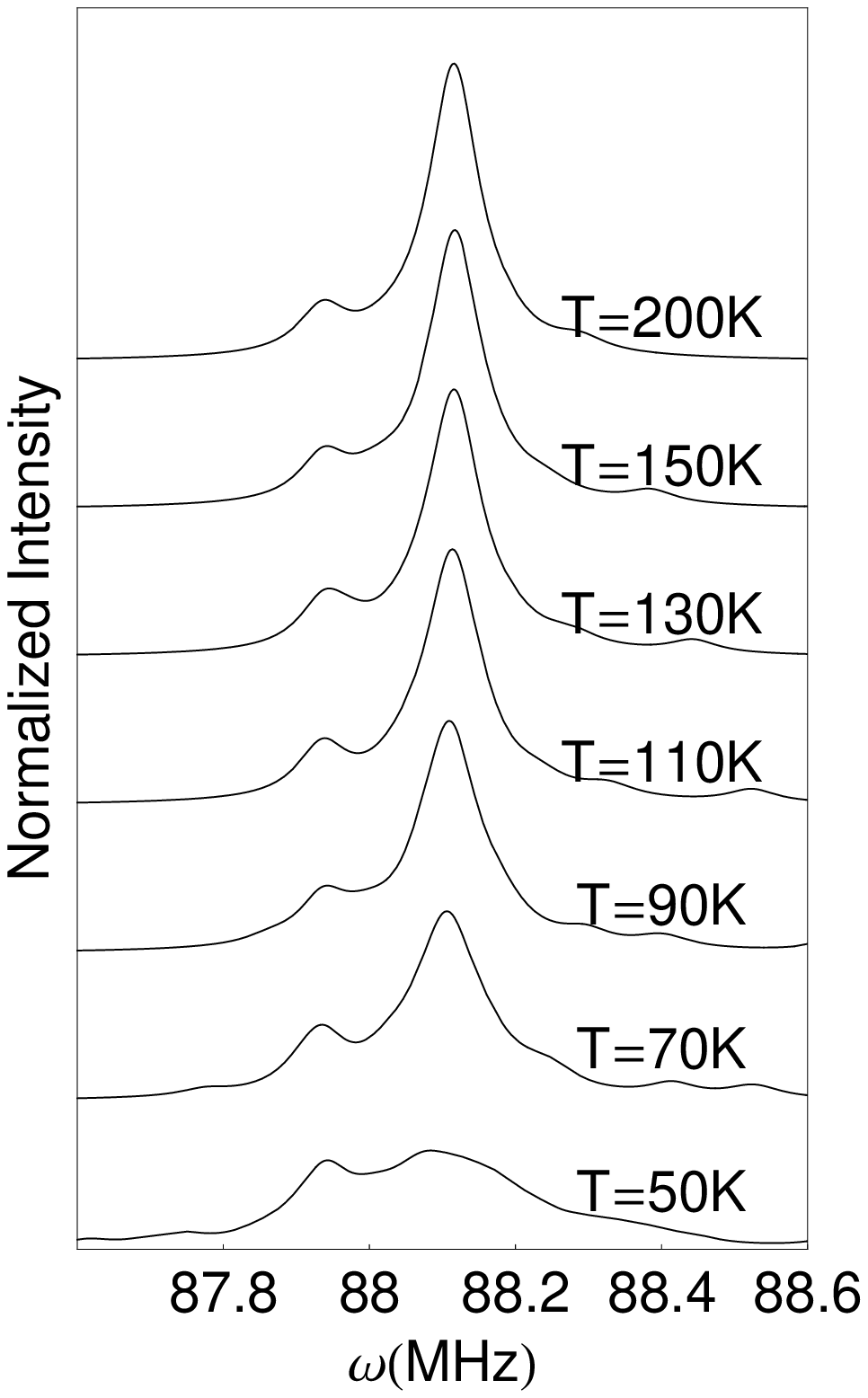}
\end{minipage}
\begin{minipage}{3.5cm}
\includegraphics[clip=true,width=3.5cm]{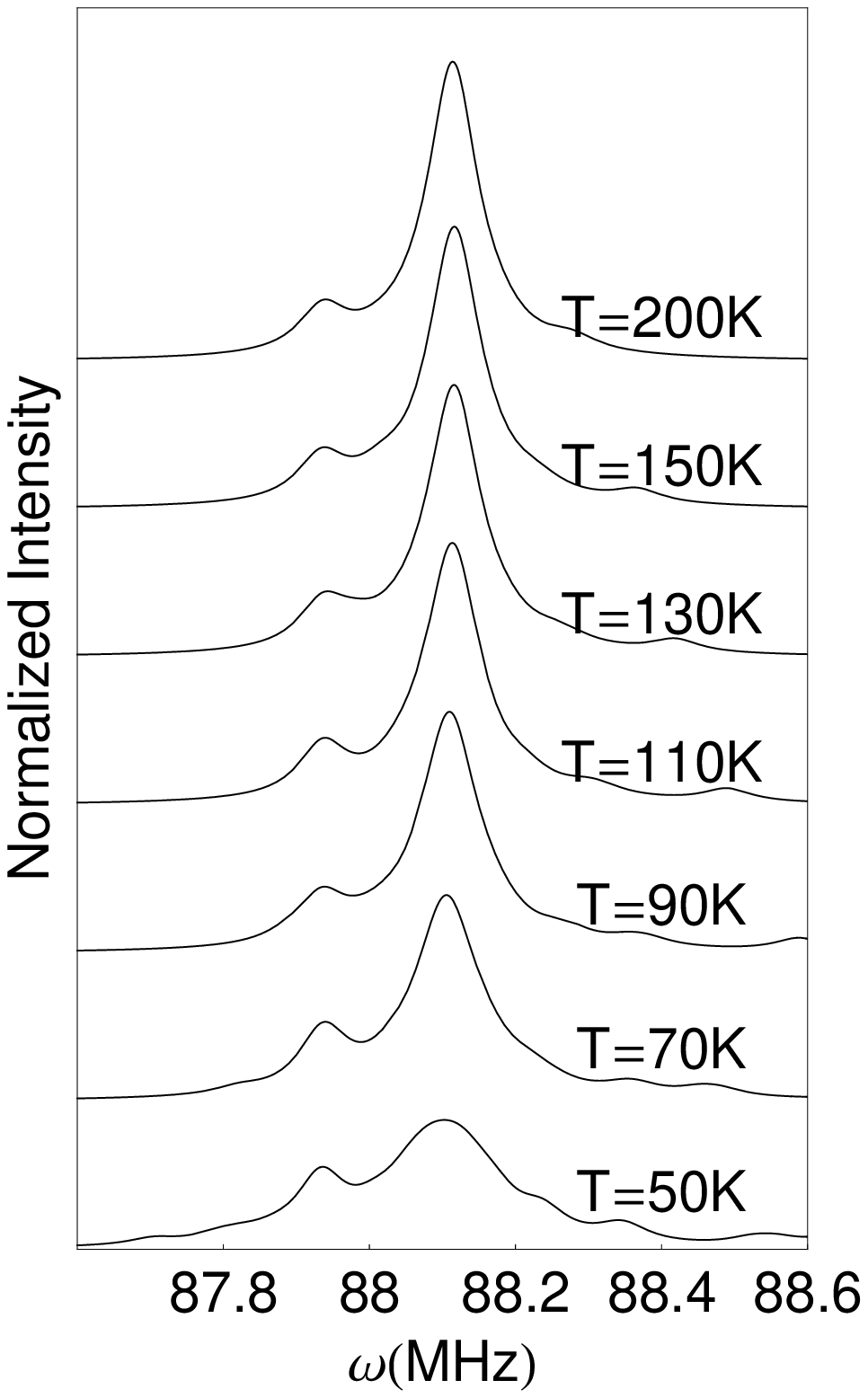}
\end{minipage}
\caption{(color online) Comparison of experimental $^{63}$Cu(1F)
NMR line of Ortho-II YBCO6.5 at external field 7.7T along ${\bf
a}$ direction(upper left), with three models studied: isolated 1D
chain with random impurities $H_{chain}+H^{(1)}_{imp}$(upper
right), isolated 1D chain with a section of consecutive Cu erased
$H_{chain}+H^{(2)}_{imp}$(lower left), chain-plane coupled system
with a section of chain Cu erased
$H_{plane}+H_{chain}+H_{inter}+H^{(2)}_{imp}$ (lower right). }
\label{fig:compare_NMR}
\end{center}
\end{figure}

\begin{figure}[t]
\begin{center}
\includegraphics[clip=true,width=0.8\columnwidth]{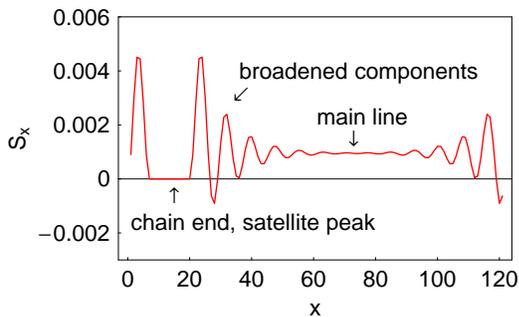}

\caption{(color online).  Real space magnetization on one of the
chains in the Cu(1F) model
$H_{plane}+H_{chain}+H_{inter}+H^{(2)}_{imp}$, at temperature
$70K$. The spinless sites, here assumed to be near the chain end,
give  zero magnetization and correspond to the satellite peak on
the NMR spectrum. The small magnetization region away from the
chain end corresponds to the main line, while the large
oscillation sector close to the spinless sites broadens the main
line. } \label{fig:eliminate_Cu_magnetization}
\end{center}
\end{figure}

The effect of removing these Cu spins is first examined in a
single isolated chain, where the elimination of $\delta t^{r}$ in
Eq. (\ref{chain_section_imp_H}) is omitted since
%
all couplings to the plane are ignored in this approximation. In
comparison with randomly distributed missing oxygens
$H_{chain}+H^{(1)}_{imp}$, Cu(1F) lines given by considering
$H_{chain}+H^{(2)}_{imp}$ show a clear asymmetric line shape with
significantly more weight at lower frequency, as shown in Fig.
\ref{fig:compare_NMR}. The zero magnetization peak appears at all
temperatures, and is specially noticeable at high temperature as
the main line narrows. Broadening of the main line smears out the
satellite peak, which may explain its apparent disappearnce at low
temperature in the work of Yamani et al.\cite{Yamani06}. A typical
real space magnetization pattern of the eliminated Cu(1F) model is
shown in Fig. \ref{fig:eliminate_Cu_magnetization}, where one can
clearly identify each sector of the magnetization with the corresponding
features in the NMR spectrum.

\section{Effect of Chain-Plane Coupling}
The conclusion from examining the single isolated chain does not
alter as the chains are coupled to the plane; in other words, we
still see (1)broadening of spectrum at low temperature,
(2)asymmetric spectral weight, and (3)the high temperature
satellite peak, as shown in Fig. \ref{fig:compare_NMR}. The
interlayer coupling   influences the pairing and magnetic state of
the  plane, as we show below, but as far as the chain
magnetization is concerned, the interlayer coupling does not
qualitatively affect its magnitude or spatial distribution.

Our basic finding here is that the chain magnetization, which has
a strongly 1D character, imprints itself on the magnetic response
of the plane to which it is coupled, even though the coupling
itself is only via interplanar hopping (there are no explicit
chain-plane exchange interactions included in the Hamiltonian).
Disorder in the plane will of course produce a response with local
2D (4fold) symmetry, as in e.g. the occasional Cu vacancy
which will produce a unitary scatterer.  We neglect these effects
in the current model, and focus only on the chain disorder which
is expected to dominate in the ultraclean Ortho-II crystals.
%
%

\begin{figure}[ht]
\centering

a)
\epsfig{file=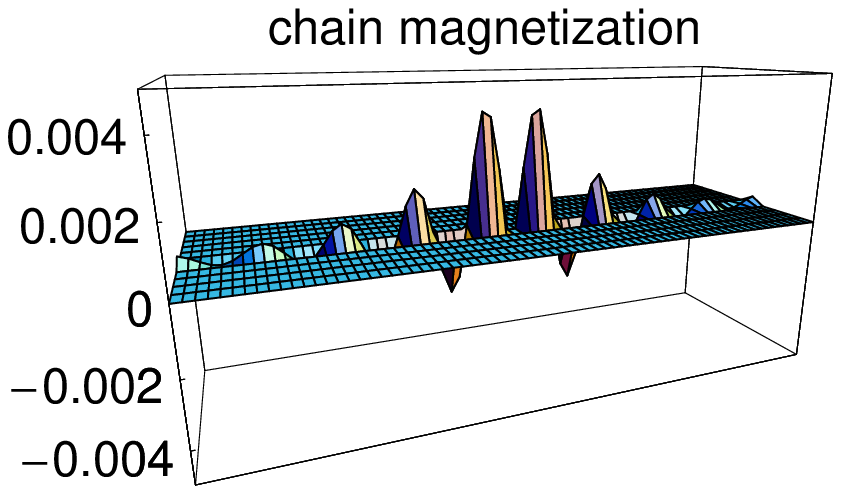,width=0.8\columnwidth,clip=true}\\
b)
\epsfig{file=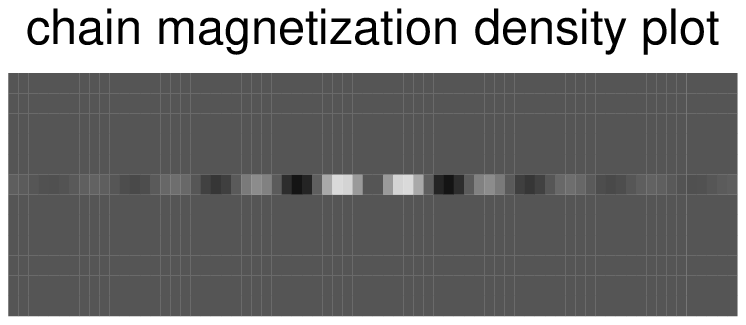,width=0.8\columnwidth,clip=true}\\
c)  \epsfig{file=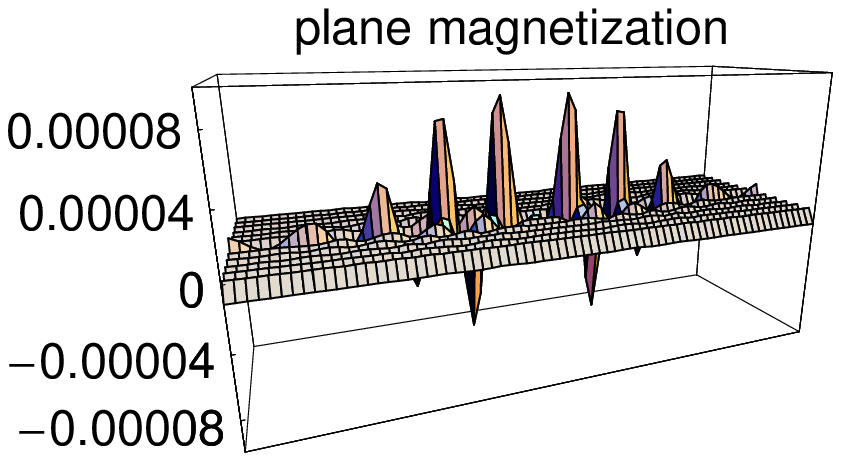,width=0.8\columnwidth,clip=true} \\

d)
\epsfig{file=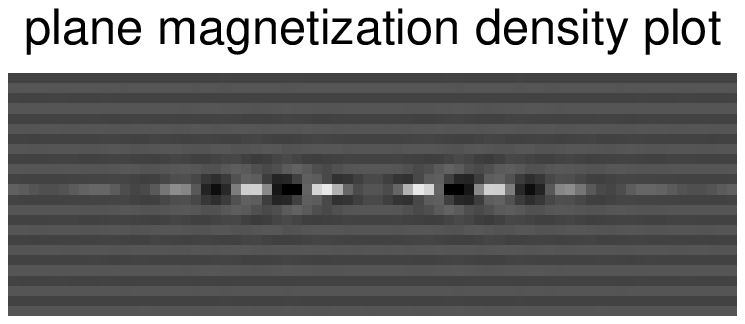,width=0.8\columnwidth,clip=true}
\caption{(color online) Magnetization induced (a)on the chain layer and (b)on the
plane by a hopping reduction defect $H^{(1)}_{imp}$ located on a single chain, together
with their corresponding false color plots (c) and (d) which
manifest the similarity between their periodicity. Size of the
plane is $72\times 24$, at temperature $70K$.  }
\label{fig:bond_on_chain}
\end{figure}

 To demonstrate this pronounced planar 1D anisotropy, as well
as to compare with the planar unitary scatterers, we employed again the
reduction of a single hopping on the chain $H^{(1)}_{imp}$ as discussed in Section II,
but this time consider the full chain-plane system $H_{plane}+H_{chain}+H_{inter}+H^{(1)}_{imp}$. Fig. \ref{fig:bond_on_chain} shows the resulting magnetization pattern
induced on both chain layer and the plane, where one again sees the long wave length oscillation
associated with the small Fermi momentum of the chain band, similar to the result of Fig.
\ref{fig:bondimp_scanU_50K_NMR}. In addition, the proximity between the two layers inherits
this small chain Fermi momentum to the plane, results in a small planar induced magnetization
that has wave length similar to that of the chain. The planar magnetization is obviously
directed along the chain direction, which confirms our suspicion that out-of-plane oxygen disorder in YBCO can induce anisotropic density and magnetic modulations on the plane.
On the other hand, numerics show that if one places a unitary scatter on the plane,
then the anisotropy of its induced magnetization is much smaller even when interlayer coupling is present. Such a hierarchy unambiguously proves that, when systems with different dimensions are coupled,
the response of lower dimension can drive that of the higher dimension, consistent
with our expectation that correlation effect becomes more significant as dimensionality is lowered.

\begin{figure}[ht]
\centering
\begin{tabular}{cccc}
a) & \epsfig{file=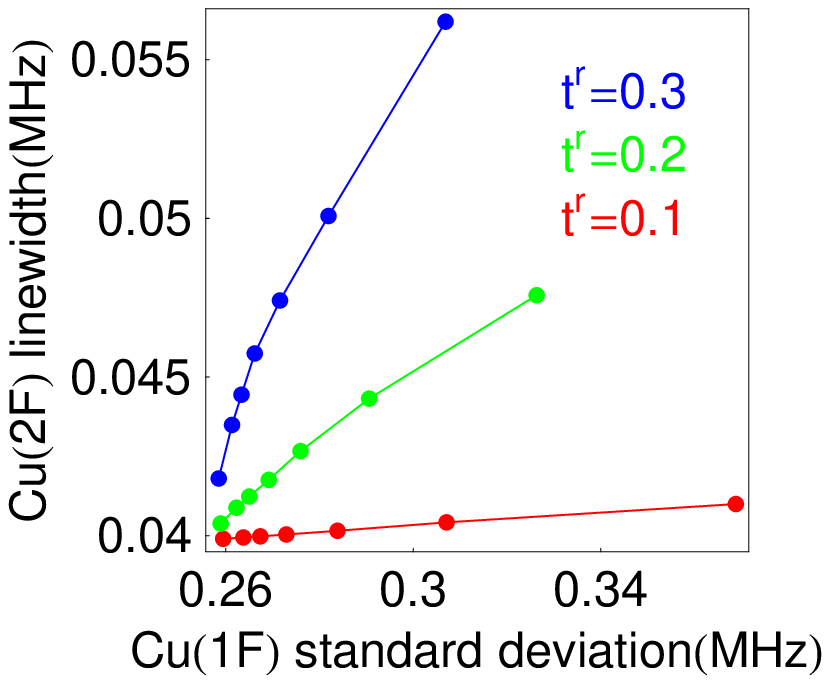,width=4cm,height=3.5cm,clip=true}
b) & \epsfig{file=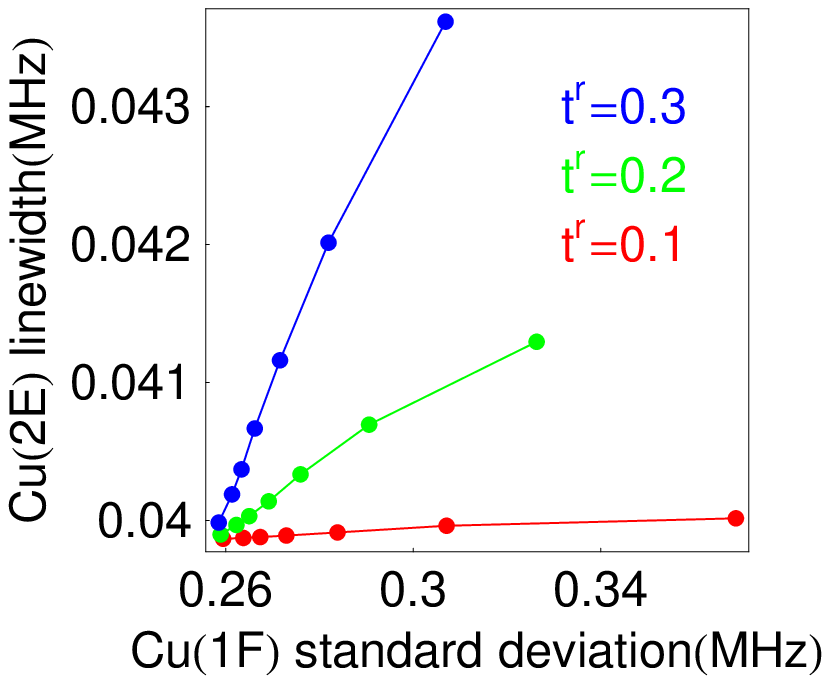,width=4cm,height=3.5cm,clip=true}
\end{tabular}
\caption{(color online) The standard deviation of Cu(1F) line versus the Lorentzian linewidth
of the (a)Cu(2F) and (b)Cu(2E) line, extracted from the chain-plane model $H_{plane}+H_{chain}+H_{inter}+H^{(2)}_{imp}$. Note that due to the high temperature satellite peak and the complex line shape of Cu(1F), as shown in Fig. \ref{fig:compare_NMR}, we use the standard deviation of the Cu(1F) line to represent its linewidth. Data taken at system size $101\times 10$ with eliminated
sites $L=11$ in impurity model Eq. (\ref{chain_section_imp_H}), and the temperature of each
point corresponds to the same scale in Fig. \ref{fig:compare_NMR}. }
\label{fig:1F2F2E_NMR_lines}
\end{figure}

Finally, we address the issue of correlation between broadening of
Cu(1F) and Cu(2E/F) lines. Experimentally the identification of
each line on the complete NMR spectrum is made via comparison with
YBCO6 and YBCO7 spectra, and the Knight shift can be extracted by
subtracting the orbital contribution $K_{orb}$ associated with
each copper species.\cite{Yamani03} The reverse process provides
the recipe of recovering the NMR spectrum in the present model,
with each line calculated by collection of magnetization and Eq.
(\ref{chain_NMR_formula}). The value of $K_{orb}$ at external
field $B//{\bf b}$ is taken from Takigawa {\em et
al}.\cite{Takigawa89} In Fig. \ref{fig:1F2F2E_NMR_lines}, we show
that the broadening of extracted Cu(2F) and Cu(2E) lines as
temperature is lowered follows that of Cu(1F), with the slope of
Cu(2E) smaller than Cu(2F). Increasing the interlayer hopping
increases the Cu(1F) and Cu(1E) linewidth, indicating that the
magnetization on the plane comes from its proximity to the chains.
Within the range of interlayer hopping that retains the Fermi
surface topology, the linewidth correlation between the planar Cu
and the chain Cu is roughly linear, similar to the conclusion
found experimentally.\cite{Yamani06}

\section{Conclusions}

In summary, we have shown that the broadening of Cu(1F) NMR lines
of ortho-II YBCO in the superconducting state at low temperatures
cannot be explained unless one assumes the existence of
intermediate strength magnetic correlations on the chains.  We
introduced a model with correlations described by Hubbard
interactions $U$ on both the plane and chains, together with a
$d$-wave pairing interaction, both of which were treated in an
unrestricted mean field approximation. In the presence of an
impurity, here assumed to be an oxygen vacancy in the nearly full
chain, we find that a staggered incommensurate magnetization
oscillation
 grows as
temperature is lowered (see Ref. \onlinecite{Alloul07}); many such
defects give rise to a broadening of NMR lines which grows as
temperature is decreased, in quantitative agreement with recent
experiments by Yamani et al.\cite{Yamani03,Yamani06}  The
Friedel-type spin density wave induced by a uncorrelated metallic
host is, by contrast, too small and temperature independent and
therefore cannot account for the broadening of NMR lines. We
further proposed an ad hoc model which may account for the high
temperature satellite peaks in the NMR spectrum observed in Refs.
\onlinecite{Yamani03,Yamani06}, by assuming a set of a few percent
Cu(1F) 3+ ions in the chains. The essential point is that these
exceptional Cu's should produce zero polarization; this then
accounts  quite well for the weight and temperature independence
of the observed satellite peaks.

Our model assumes pairing interactions in the plane alone, and
superconductivity is induced in the chains only by ``proximity
coupling", i.e. a chain-plane electron hopping.  This coupling is
then found to cause an imprint of the defect-induced magnetic
chain correlations on the plane itself, consisting of an
incommensurate spin density wave of quasi-1D character.   This
signal is an order of magnitude smaller than the chain
polarization for realistic parameters but should still be
observable. Indeed, within our model the broadening of planar NMR
lines is proportional to that of the chain NMR line, consistent
with the observation by Yamani et al.\cite{Yamani03,Yamani06}
These defect-induced 1D correlations should enhance the effect of
the chain bands on the magnetic response of the YBCO system at low
energies, particularly for low dopings near the Ortho-II and III
oxygen concentrations where an integer number of chains are
filled. Further research into various aspects of this response is
in progress.

{\it Acknowledgments.}

We  gratefully acknowledge useful discussions with Z. Yamani, W. A. Atkinson,
D.A. Bonn, J. Bobroff, J.P. Carbotte, P. Dai, M. Gabay,  W.N.
Hardy, Y. Sidis and O.P. Sushkov.


\newpage

\end{document}